\documentclass[review,12pt]{elsarticle}
\usepackage[utf8]{inputenc}
\usepackage{graphicx}
\usepackage{subcaption}
\usepackage{hyperref}
\usepackage{amssymb}
\usepackage{amsthm}
\usepackage{amsmath}
\biboptions{longnamesfirst,sort&compress}
\usepackage{xcolor}
\usepackage[version=4]{mhchem}

\journal{}
\begin{document}
	
	\begin{frontmatter}
		\title{Single-morphogen Turing instability driven by nonlinear intracellular–extracellular coupling}
		
		\author[mymainaddress]{Alejandro Valdés López}
		\author[mysecondaryaddress]{D. Hern\'andez}
		\author[mymainaddress]{E. C. Herrera-Hern\'andez\corref{mycorrespondingauthor}}
		\cortext[mycorrespondingauthor]{Corresponding author}
		\ead{erik.herrera@uaslp.mx}
		\address[mymainaddress]{Centro de Investigación y Estudios de Posgrado, Facultad de Ciencias Químicas, Universidad Autónoma de San Luis Potosí, Av. Dr. Manuel Nava 6, Zona Universitaria, 78210, San Luis Potosí, México}
		\address[mysecondaryaddress]{Posgrado en Ciencias de la Complejidad. Universidad Autónoma de la Ciudad de México. San Lorenzo 290, esquina Roberto Gayol, Col. del Valle Sur, Benito Juárez, CDMX, 03100, México}

		\begin{abstract}
			We show that compartmentalizing a single molecular species into intracellular and extracellular fields, and coupling them by means of membrane transport or nonlinear basal production rates, can produce diffusion-driven (Turing) instabilities. By linearizing the two-field system we derive the corresponding Turing conditions under which such instabilities can emerge. We present three biologically motivated examples that satisfies these conditions and demonstrate the resulting spatial patterns through numerical simulations. These results indicate that tissue compartmentalization alone might enable pattern formation traditionally attributed to multi-species systems.
		\end{abstract}

		
		

		
	\end{frontmatter}
	
	
	\section{Introduction}
	\label{sec:introduction}

	Spontaneous emergence of spatial order from an initially homogeneous distribution of chemical species is one of the central problems of developmental biology. Alan Turing's seminal work \cite{turing_chemical_1952} established that a homogeneous steady state, stable in the absence of diffusion, can become unstable when two reacting and diffusing chemical species—morphogens—interact under specific contitions. This activator–inhibitor logic was later formalized and popularized by Gierer and Meinhardt \cite{gierer_theory_1972} and by Murray \cite{murray_mathematical_1989}, who emphasized the need for local self-activation coupled with longer-range antagonism as the minimal motif for robust patterning. This mechanism provides a plausible and widely invoked explanation for the generation of periodic spatial patterns, including pigmentation stripes and spots \cite{kondo_reaction-diffusion_2010, Kondo2021Studies}, digit spacing during limb development \cite{raspopovic_digit_2014}, and hair follicle spacing \cite{mooney_spatial_1985,sick_wnt_2006}. However, it has also been criticized because it imposes strict requirements on the underlying chemical network, motivating numerous extensions of the canonical two-morphogen architecture, seeking to reconcile molecular realism with pattern-forming requirements \cite{landge_pattern_2020,paul_widespread_2024}.

	Mathematical analysis shows that a single diffusing morphogen governed by a local reaction–diffusion equation with zero–flux (homogeneous Neumann) boundary conditions cannot exhibit a classical Turing instability, irrespective of the complexity of its reaction kinetics \cite{szili_origin_1997, mendez_reactiontransport_2010} — or even if anomalous or fractional diffusion operators are considered \cite{henry_fractional_2000}. After linearization around a homogeneous steady state, the growth rate takes the form $\lambda(k) = f'(u^*) - D_u\phi(k)$, where $\phi(k) \ge 0$ for both classical and fractional diffusion. Hence, if the steady state is locally stable $f'(u^*) < 0$, all spatial modes remain stable.

	The impossibility of diffusion-driven destabilization in one-component systems is therefore a structural consequence of the scalar spectral problem, which lacks the activator–inhibitor coupling required to separate local stability from spatial instability \cite{satnoianu_turing_2000}. This structural limitation has recently motivated several approaches that exploit spatial heterogeneity rather than chemical diversity to bypass the scalar constraint. Single-morphogen patterns have been reported in cell lattice models \cite{wang_periodic_2022}, mechanochemical systems \cite{recho_theory_2019}, and, very recently, in systems of layered two-dimensional media with nonlinear interlayer coupling \cite{mahashri_patterns_2026}. The present work contributes to this emerging direction by showing that the biologically compartmentalization of tissues into intracellular and extracellular domains provides a natural and mechanistically grounded realization of the same principle.
	
	Despite the success of diffusion-driven instability models in explaining diverse biological patterning phenomena, it is important to recognized that real tissues cannot be faithfully represented as homogeneous continua. Biological media are intrinsically structured since cells partition the reaction space into chemically distinct intracellular and extracellular domains, separated by membranes that regulate molecular exchange through passive and/or active transport mechanisms \cite{keener_mathematical_2009}. This compartmentalization introduces additional layers of spatial organization and regulatory complexity absent from classical homogeneous reaction–diffusion descriptions and motivated the development of the models presented in \cite{hernandez_mean-field_2026}. In this framework each chemical species is described by two coupled scalar fields —one intracellular and one extracellular— each one evolving under its own diffusivity and reaction environment, and interacting through membrane transport. 
	
	Motivated by this structure, we pose the question: can a single chemical species support the formation of Turing patterns under these conditions? Next, we show that this structural doubling is sufficient to enable Turing instability for a single molecular species, provided that the membrane transport or the basal production rate of the morphogen coupling both fields is nonlinear in a way that implements effective activator–inhibitor feedback. The general linear stability conditions are derived, and numerical evidence of Turing patterns arising from a biologically motivated examples are presented—demonstrating that compartmentalization can unlock the pattern-forming capacity classically attributed only to multi-morphogen systems.
	
	\section{Minimal mechanism to generate Turing patterns with a single morphogen}
	\label{sec:analysis}

	We work within the bi-percolating framework of \cite{hernandez_mean-field_2026}, in which a single molecular species $U$ is described by its intracellular and extracellular concentrations. The elementary kinetic mechanism is as follows:

	\begin{equation}
		\begin{aligned}
			&\textit{Intracellular medium:} & & \ce{\text{Cell} ->[k_1] U_i}, \quad  \ce{U_i ->[k_2] D},\\
			&\textit{Membrane transport:} & & \ce{U_i <-->[\Gamma_u][\Gamma_u] U_o},\\
			&\textit{Extracellular medium:} & & \ce{U_o ->[k_3] D}.\\
		\end{aligned}\label{ec:minimal_mechanism}
	\end{equation}

	The reactions correspond to the production of the molecule $U$ by the cell, degradation/transformation into the molecule $D$, and a transport in the cell membrane that couples the intracellular and extracellular compartments.\emph{The degradation steps are structurally necessary, otherwise the Jacobian of the homogeneous system is $\det(J) = 0$ at any steady state, preventing the Turing conditions from being satisfied}. The corresponding reaction–diffusion system according to the original derived framework is:
	
	\begin{equation}
		\begin{aligned}
			\frac{\partial u_i}{\partial t} &= D_{u,i} \nabla^2u_i + \alpha_u + \beta_{u,u} u_i - \frac{\rho_c s_c }{\Delta l} \Lambda_u \\
			\frac{\partial u_o}{\partial t} &= D_{u,o} \nabla^2 u_o - \gamma_u u_o + \frac{\rho_c s_c }{\Delta l} \Lambda_u
		\end{aligned}\label{ec:flux_plus_degradation}
	\end{equation}
	where $\alpha_u = \frac{\rho_c}{k_1}$, $\beta_{u,u}=-k_2$, ${\gamma}_u=-k_3$, and $\Lambda_u$ is the net membrane flux (positive from intracellular to extracellular). If passive transport is considered, the expression for $\Lambda_u$ is given by:
	
	\begin{equation}\label{ec:pas}
		\begin{split}
			\Lambda_{pa}=\Lambda_{nd}+\Lambda_{nld}+\Lambda_{cd}+\Lambda_{ad},\\
			\Lambda_{nd}=-D_{u,m}\nabla u,\\
			\Lambda_{nld}=-D_{u,m}\left(u\right)\nabla u,\\
			\Lambda_{cd}=-\sum_{k}c_{u,k}(u,k)\nabla k, k\neq u,\\
			\Lambda_{ad}=-c_{NF,J}\Biggl[Notice \int_0^1d\xi\phi(\xi)\frac{\partial^{1-\xi}}{\partial t^{1-\xi}}\Biggr] \nabla u-c_{MF,J}\int_0^t\Xi(t-t')\nabla u dt',\\
			\Xi(t)=L^{-1}\Biggl( \biggl[\int_0^1\hat{\phi}(\xi)s^{\xi-1}d\xi\biggr]^{-1}\Biggr) 0<\xi\leq 1, \quad k\in \{A,B,...,M\}.
		\end{split}
	\end{equation}
	where the expressions represent the contributions of normal, nonlinear, cross and anomalous transport.  Notice that $\Gamma_u = \frac{\rho_c s_c D_{u,m}}{\Delta l}$ acts as an effective membrane exchange coefficient (with units of $T^{-1}$) and incorporates tissue-specific properties—namely, cell surface area, average cell density in the tissue, the diffusion coefficient of the species within the cellular membrane, and membrane thickness—thereby ensuring dimensional consistency.
	
	For system \eqref{ec:flux_plus_degradation}, the dispersion relation is computed as the real part of the largest eigenvalue ($\text{max} \, \Re (\lambda(k))$) of the matrix shown in Eq.~\eqref{ec:jacobian_general_mechanism}, while for a Turing pattern to grow, the conditions given in Eq.~\eqref{ec: general_conditions_one_morphogen} must be satisfied, as explained in \cite{woolley_bespoke_2021}.
	
	\begin{equation}
		J =\left(\begin{array}{cc} -\Gamma _{u}\frac{\partial \Lambda _{u}}{\partial u_{i}} -k_{2} -D_{u,i}k^2& -\Gamma _{u}\frac{\partial \Lambda _{u}}{\partial u_{o}} \\ \Gamma _{u}\frac{\partial \Lambda _{u}}{\partial u_{i}}  & \Gamma _{u}\frac{\partial \Lambda _{u}}{\partial u_{o}} -k_{3}-D_{u,o}k^2 \end{array}\right) \label{ec:jacobian_general_mechanism}
	\end{equation}
	
	\begin{equation}
		\begin{aligned}
			&\operatorname{tr}(J)<0, 
			\qquad 
			\det(J)>0, \\
			&-D_{u,o}\left(\Gamma _{u}\frac{\partial \Lambda _{u}}{\partial u_{i}}+k_{2}\right)
			-D_{u,i}\left(k_{3}-\Gamma _{u}\frac{\partial \Lambda _{u}}{\partial u_{o}}\right)>0, \\
			&\left(D_{u,o}\left(\Gamma _{u}\frac{\partial \Lambda _{u}}{\partial u_{i}}+k_{2}\right)
			+D_{u,i}\left(k_{3}-\Gamma _{u}\frac{\partial \Lambda _{u}}{\partial u_{o}}\right)\right)^2
			>4D_{u,i}D_{u,o}\det(J), \\
			&\operatorname{tr}(J)=
			-\Gamma _{u}\frac{\partial \Lambda _{u}}{\partial u_{i}}
			+\Gamma _{u}\frac{\partial \Lambda _{u}}{\partial u_{o}}
			-k_{2}-k_{3}, \\
			&\det(J)=
			\Gamma _{u}k_{3}\frac{\partial \Lambda _{u}}{\partial u_{i}}
			-\Gamma _{u}k_{2}\frac{\partial \Lambda _{u}}{\partial u_{o}}
			+k_{2}k_{3}.
		\end{aligned}\label{ec: general_conditions_one_morphogen}
	\end{equation}

	Observing Eq.~\eqref{ec:jacobian_general_mechanism}, since $\Gamma_u$ must be strictly positive—as all parameters composing it are physical characteristics of the cell—for $J_{11}$ to be positive in the absence of diffusion it is necessary that $\frac{\partial \Lambda_u}{\partial u_i} < 0$, with this derivative large enough in magnitude to overcome the stabilizing contribution of $k_2$. This condition implies that the only admissible sign structure for the Jacobian is of the \emph{cross activator–inhibitor} type \cite{mendez_reactiontransport_2010}, $J = \begin{pmatrix} + & + \\ - & - \end{pmatrix}$, for which $J_{12} > 0$ further requires $\frac{\partial \Lambda_u}{\partial u_o} < 0$. Together, these two conditions impose special requirements on the membrane flux function $\Lambda_u(u_i, u_o)$: an increase in $u_i$ inhibits $u_o$ (via $J_{21} < 0$), while an increase in $u_o$ activates $u_i$ (via $J_{12} > 0$), generating a coupling in which both concentrations of morphogen $u$ tend to vary together, producing in-phase spatial profiles in which peaks coincide \cite{murray_mathematical_1989}.

	To show that the Turing instability conditions derived in Eq.~\eqref{ec: general_conditions_one_morphogen} can be realized by a physically motivated transport law, we proceed guided by three criteria: (i) analytical verifiability of the instability conditions, (ii) biological plausibility, and (iii) the ability to produce spatial patterns. The third requirement is non-trivial, since satisfying the linear instability conditions alone is necessary but not sufficient for pattern formation \cite{murray_mathematical_1989}. As an illustrative realization, we consider two forms of concentration-dependent nonlinear membrane transport, given by Eq.~\eqref{ec:TP_flux}.
	
	\begin{equation}
		\begin{aligned}
			\Lambda_u \left( u_i,u_o \right) &= D_{u,m} \left( u_i - u_o \right) \\[6pt]
			D_{u,m} &= D_{u,m}^0 \begin{cases}
				\dfrac{K_i^n}{K_i^n + u_i^n}\cdot \dfrac{u_o^m}{K_o^m + u_o^m} & \text{Case 1} \\[10pt]
				\dfrac{K^n}{K^n + \left( u_i + u_o \right)^n} & \text{Case 2}
			\end{cases}
		\end{aligned}\label{ec:TP_flux}
	\end{equation}
	
	In Case 1, the effective permeability $D_{u,m}$ is governed by two opposing regulatory factors: a decreasing Hill function in $u_i$, so that higher intracellular concentration suppresses permeability (with $K_i$ as the half-inhibition constant and $n$ as the cooperativity exponent), and an increasing Hill function in $u_o$, so that higher extracellular concentration enhances permeability (with $K_o$ and $m$ as the corresponding parameters). This asymmetric dependence is intended to implement an activator–inhibitor-like feedback across the membrane: the extracellular morphogen enhances permeability while the intracellular concentration limits it, introducing a nontrivial coupling between driving force and effective transport rate. In Case 2, $D_{u,m}$ follows a symmetric saturating Hill function of the total concentration $u_i + u_o$, so that permeability decreases as the combined morphogen load increases. Here $K$ sets the half-saturation threshold and $n$ controls the cooperativity of the response, making this case appropriate for modeling non-selective membrane crowding or competitive occupancy of transport sites.

	In both cases, since $D_{u,m}$ is a decreasing function of $u_i$ while $(u_i-u_o)$ is increasing in $u_i$, the sign of $\frac{\partial \Lambda_u}{\partial u_i}$ is not fixed \emph{a priori} but depends on which effect dominates at the steady state. Therefore, the Turing conditions are satisfied only in specific parameter regimes, as verified numerically below. Although neither case is intended as a specific molecular model, such concentration-dependent permeability can arise when the local chemical environment modifies membrane transport pathways — for example, through conformational or structural changes in transport proteins or membrane pores \cite{gudyka_concentration-dependent_2024, queralt-martin_ion_2026}. Similar nonlinear force–flux relations have recently been formalized in theoretical descriptions of chemo-responsive membranes with feedback-controlled permeability \cite{kim_permeability_2022,milster_feedback-controlled_2023}.
	
\begin{figure}[htbp]
		\centering
		\begin{subfigure}{0.5\linewidth}
			\centering
			\includegraphics[width=\linewidth]{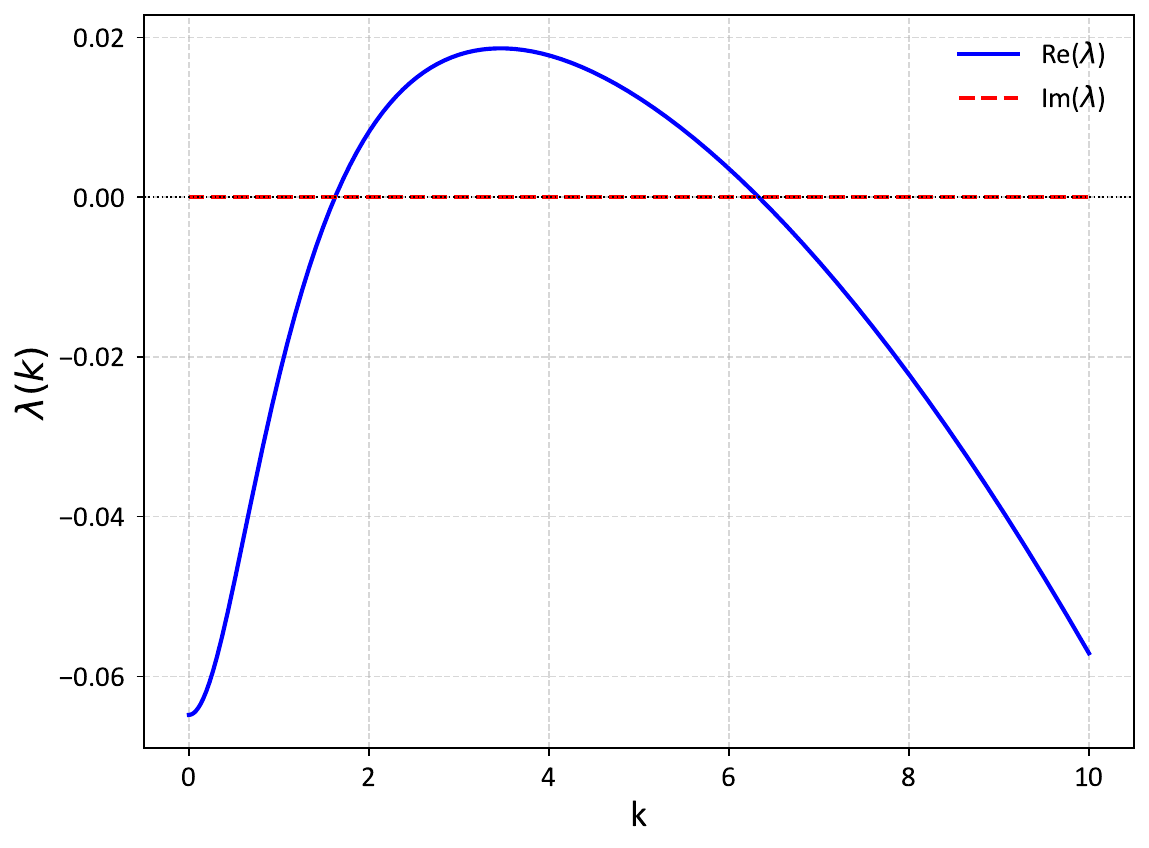}
			\caption{Dispersion relation, Case 1.}
			\label{fig:DR_case1}
		\end{subfigure}
		\hfill
		\begin{subfigure}{0.4\linewidth}
			\centering
			\includegraphics[width=\linewidth]{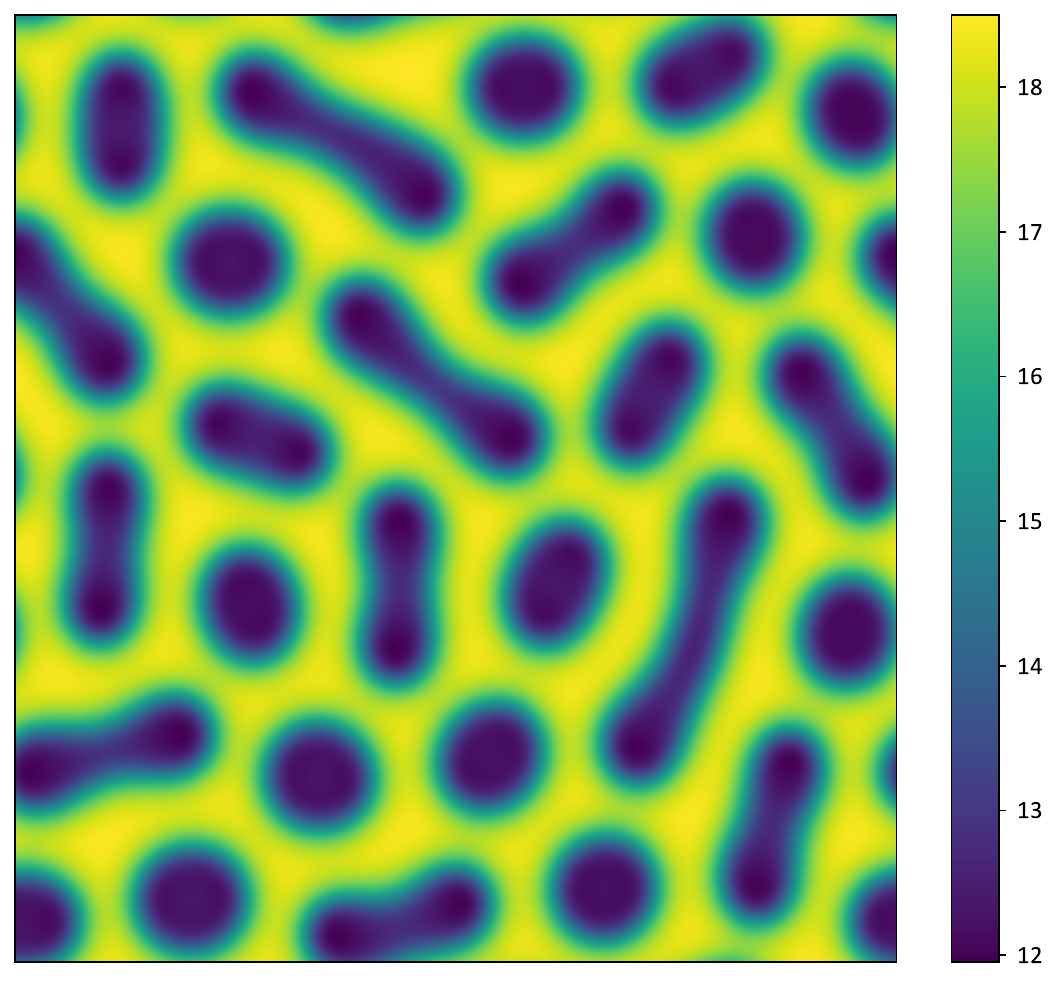}
			\caption{Turing pattern, Case 1.}
			\label{fig:TP_case1}
		\end{subfigure}
		
		\vspace{1.5em}
		
		\begin{subfigure}{0.5\linewidth}
			\centering
			\includegraphics[width=\linewidth]{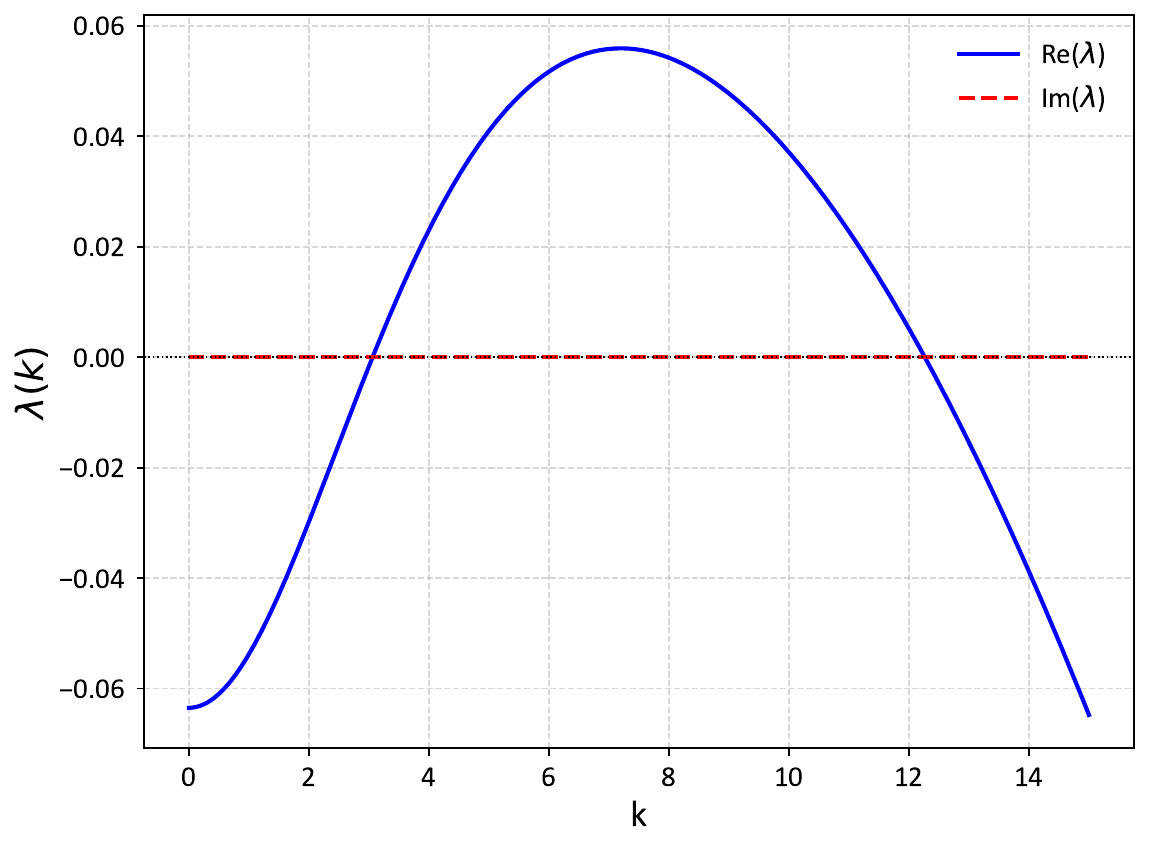}   
			\caption{Dispersion relation, Case 2.}
			\label{fig:DR_case2}
		\end{subfigure}
		\hfill
		\begin{subfigure}{0.4\linewidth}
			\centering
			\includegraphics[width=\linewidth]{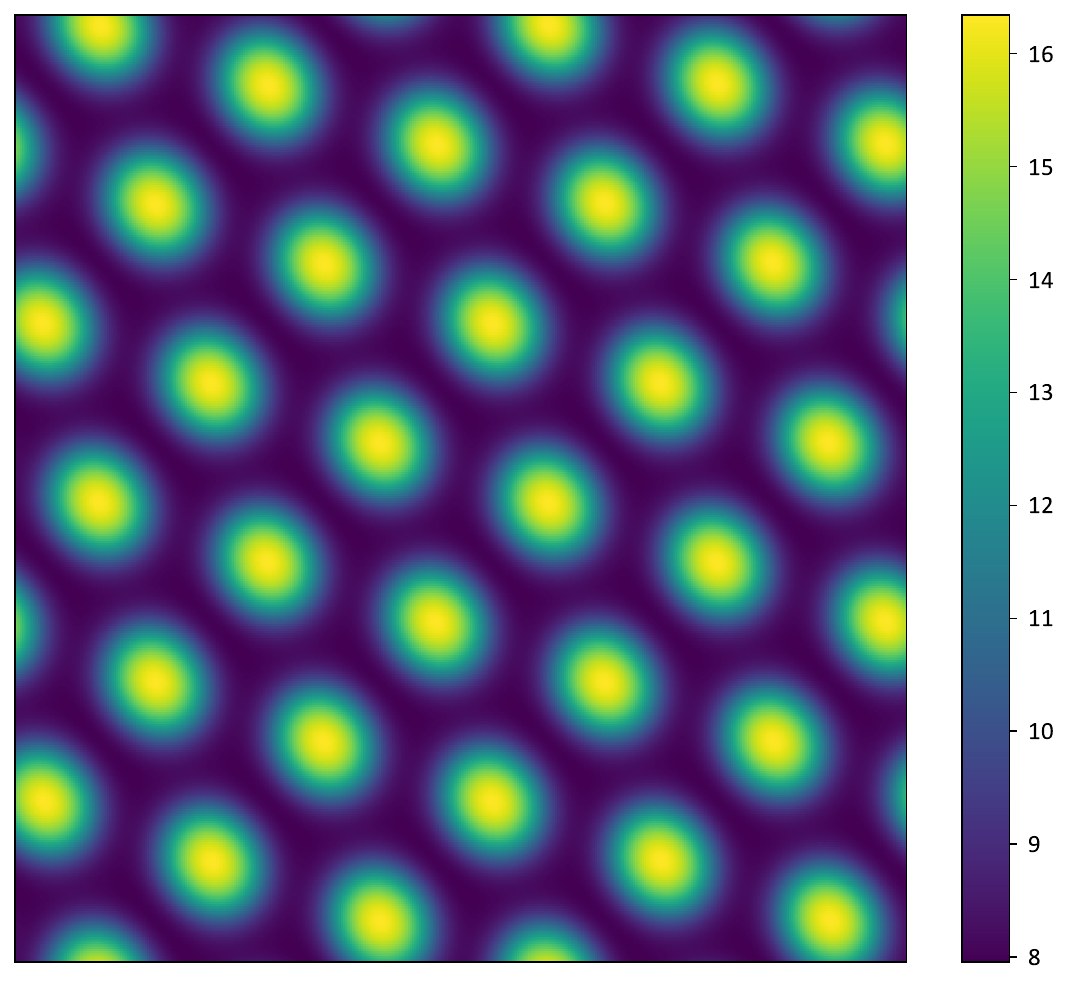}   
			\caption{Turing pattern, Case 2.}
			\label{fig:TP_case2}
		\end{subfigure}
		
		\caption{Dispersion relations (left column) and Turing patterns (right column) for
			the system~\eqref{ec:flux_plus_degradation} with the nonlinear membrane flux~\eqref{ec:TP_flux}.
			\textbf{Top row (Case 1):} activator--inhibitor-type flux with opposing Hill functions.
			Parameters: $\alpha_u = 6.26526$, $\beta_{u,u} = 0.24343$, $\gamma_u = 0.29462$,
			$K_i = 5.17348$, $K_o = 4.38237$, $m = 2$, $n = 4$,
			$\Gamma_u = 35.50578$, $D_{u,i} = 0.001$, $D_{u,o} = 0.2$,
			$u_i^* = 16.18865$, $u_o^* = 7.88968$.
			\textbf{Bottom row (Case 2):} Hill-type saturable flux.
			Parameters: $\alpha_u = 9.70927$, $\beta_{u,u} = 0.32906$, $\gamma_u = 2.46063$,
			$K = 6.00629$, $n = 3$, $\Gamma_u = 8.89131$, $D_{u,i} = 0.001$,
			$D_{u,o} = 0.1$, $u_i^* = 10.06359$, $u_o^* = 2.60004$.
			In both cases homogeneous Neumann boundary conditions are imposed; while the domain is a square of side $L = 10$ for Case~1 and $L = 5$ for Case~2.}
		\label{fig:TP_DR_TP}
	\end{figure}

	The dispersion relations and Turing patterns for both cases are shown in Figure~\ref{fig:TP_DR_TP}. Numerical simulations were performed using the \texttt{py-pde} Python library \cite{zwicker_py-pde_2020}, which provides efficient solvers for partial differential equations via the method of lines. Spatial derivatives were approximated on a uniform Cartesian grid of $N \times N$ points with spacing $\Delta x = 0.05$ length units, and time integration was carried out using an explicit Euler scheme with adaptive time stepping. Initial conditions consisted of the homogeneous steady state $(u_i^*, u_o^*)$ perturbed by spatially uncorrelated uniform random noise of amplitude $\epsilon = 10^{-5}$. Simulations were run until $T_{\mathrm{final}} = 5000$, at which point the spatial variance of the solution had reached a stationary value.

	It is important to remark that, although the two cases presented demonstrate that nonlinear membrane transport alone is sufficient to drive Turing pattern formation, not every choice of $\Lambda_u(u_i, u_o)$ satisfying the conditions of Eq.~\eqref{ec: general_conditions_one_morphogen} will necessarily produce patterns, since the underlying mechanism remains parametrically sensitive in the sense discussed in \cite{maini_turings_2012, vittadello_turing_2021}. The natural extension to broaden the parameter space compatible with pattern formation would be to incorporate auto-activation mechanisms in the intracellular environment, as reported experimentally\footnote{For example, transforming growth factor beta ($TGF-\beta$), which regulates diverse cellular functions \cite{kondo_reaction-diffusion_2010, enomoto_autocrine_2023}.}, or processes that enhance coupling between intracellular and extracellular media. \emph{We excluded these mechanisms to highlight the role of nonlinear membrane transport in generating such structures on its own.}

	Up to this point, we have assumed a constant basal intracellular production rate ($\alpha_u$); however, mechanisms such as quorum sensing indicate that the extracellular concentration of the same molecule can regulate its intracellular production \cite{waters_quorum_2005, moreno-gamez_quorum_2023}. More generally, it is well recognized that the composition of the extracellular environment strongly influences intracellular signaling dynamics \cite{kholodenko_cell-signalling_2006, purvis_encoding_2013}, as illustrated by the EGFR–MAPK cascade \cite{shvartsman_autocrine_2002}. This motivates considering a concentration-dependent basal production rate $\alpha_u(u_i, u_o)$, in which intracellular production is regulated by both compartments. In this case, even a linear membrane flux of the form $\Lambda_u = D_{u,m}(u_i - u_o)$ is sufficient to generate Turing instability, since the nonlinearity required for activator–inhibitor feedback is now encoded in the production term rather than in the transport law. The governing equations become:
	
	\begin{equation}
		\begin{aligned}
			\frac{\partial u_i}{\partial t} &= D_{u,i} \nabla^2u_i + \alpha_u (u_i,u_o) + \beta_{u,u} u_i - \Gamma_u (u_i -u_o) \\
			\frac{\partial u_o}{\partial t} &= D_{u,o} \nabla^2 u_o - \gamma_u u_o + \Gamma_u (u_i -u_o) 
		\end{aligned}\label{ec:non_linear_basal} 
	\end{equation}

The Jacobian and the corresponding Turing conditions follow from linearizing around the homogeneous steady state $(u_i^*, u_o^*)$, yielding Eqs.~\eqref{ec:basal_disp_rel}–\eqref{ec:general_conditions_basal}. Inspection of these conditions reveals that instability requires $\frac{\partial \alpha_u}{\partial u_i} > 0$ and $\frac{\partial \alpha_u}{\partial u_o} < 0$, i.e. the production rate must be self-activating with respect to the intracellular concentration and inhibited by the extracellular one. By applying an analysis analogous to that developed for the nonlinear membrane transport case, it can be shown that the only admissible sign structure for the Jacobian is the \emph{pure activator–inhibitor type} \cite{mendez_reactiontransport_2010}, given by $J = \begin{pmatrix} + & - \\ + & - \end{pmatrix}$.

\begin{equation}
	J =\left(\begin{array}{cc} \frac{\partial \alpha _{u}}{\partial u_{i}} -\mathrm{Dui}\,k^2-\Gamma _{u}-\beta _{\mathrm{uu}} & \frac{\partial \alpha _{u}}{\partial u_{o}} +\Gamma _{u}\\ \Gamma _{u} & -\mathrm{Duo}\,k^2-\Gamma _{u}-\gamma _{u} \end{array}\right)
	\label{ec:basal_disp_rel}
\end{equation}

\begin{equation}
	\begin{aligned}
		&\operatorname{tr}(J)<0,
		\qquad
		\det(J)>0,\\[4pt]
		&-D_{u,o}\left(
		-\frac{\partial \alpha_u}{\partial u_i}
		+\Gamma_u+\beta_{uu}
		\right)
		-
		D_{u,i}\left(
		\Gamma_u+\gamma_u
		\right)>0,\\[4pt]
		&
		\left(
		D_{u,o}\left(
		-\frac{\partial \alpha_u}{\partial u_i}
		+\Gamma_u+\beta_{uu}
		\right)
		+
		D_{u,i}\left(
		\Gamma_u+\gamma_u
		\right)
		\right)^2
		>
		4D_{u,i}D_{u,o}\det(J),\\[4pt]
		&\operatorname{tr}(J)=
		\frac{\partial \alpha_u}{\partial u_i}
		-2\Gamma_u
		-\beta_{uu}
		-\gamma_u,\\[4pt]
		&\det(J)=
		\Gamma_u\beta_{uu}
		+\Gamma_u\gamma_u
		+\beta_{uu}\gamma_u
		-\Gamma_u\frac{\partial \alpha_u}{\partial u_i}
		-\Gamma_u\frac{\partial \alpha_u}{\partial u_o}
		-\gamma_u\frac{\partial \alpha_u}{\partial u_i}.
	\end{aligned}
	\label{ec:general_conditions_basal}
\end{equation}

\begin{figure}[htbp]
	\centering
	\begin{subfigure}{0.5\linewidth}
		\centering
		\includegraphics[width=\linewidth]{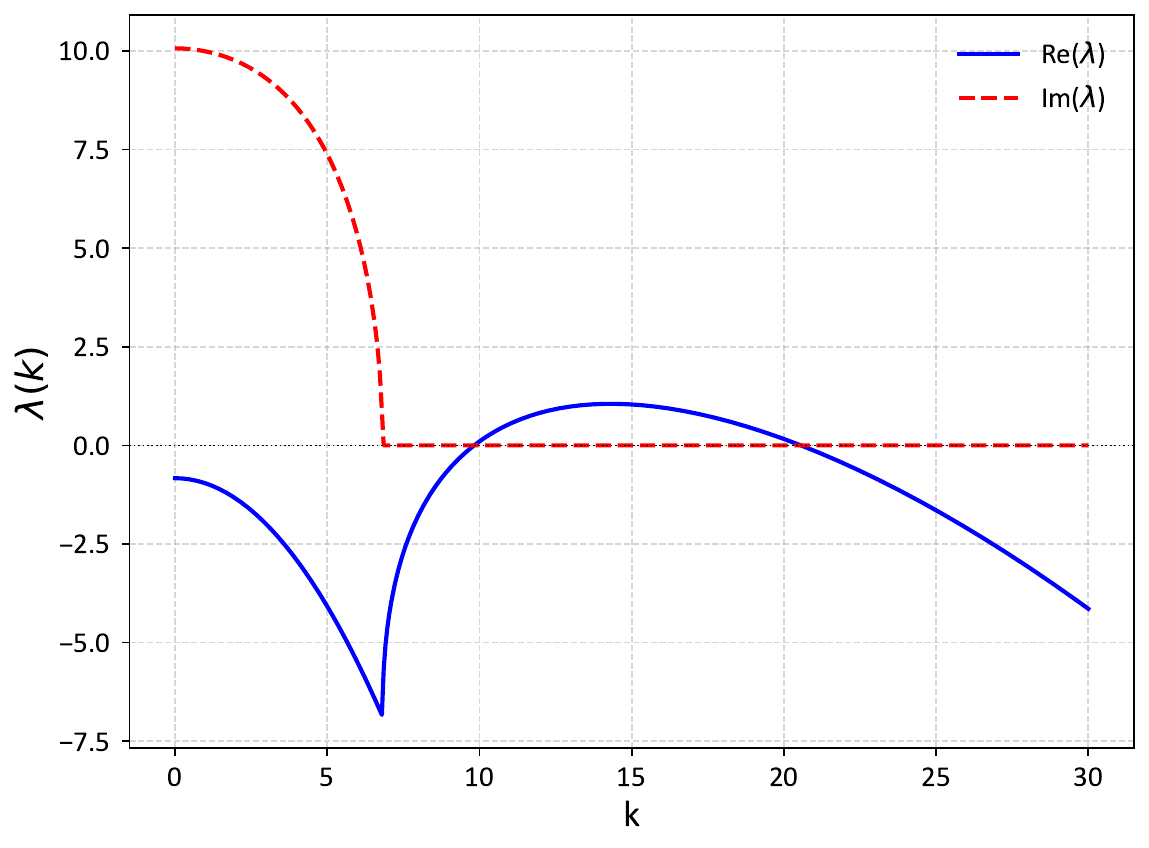}
		\caption{Dispersion relation.}
		\label{fig:DR_basal}
	\end{subfigure}
	\hfill
	\begin{subfigure}{0.4\linewidth}
		\centering
		\includegraphics[width=\linewidth]{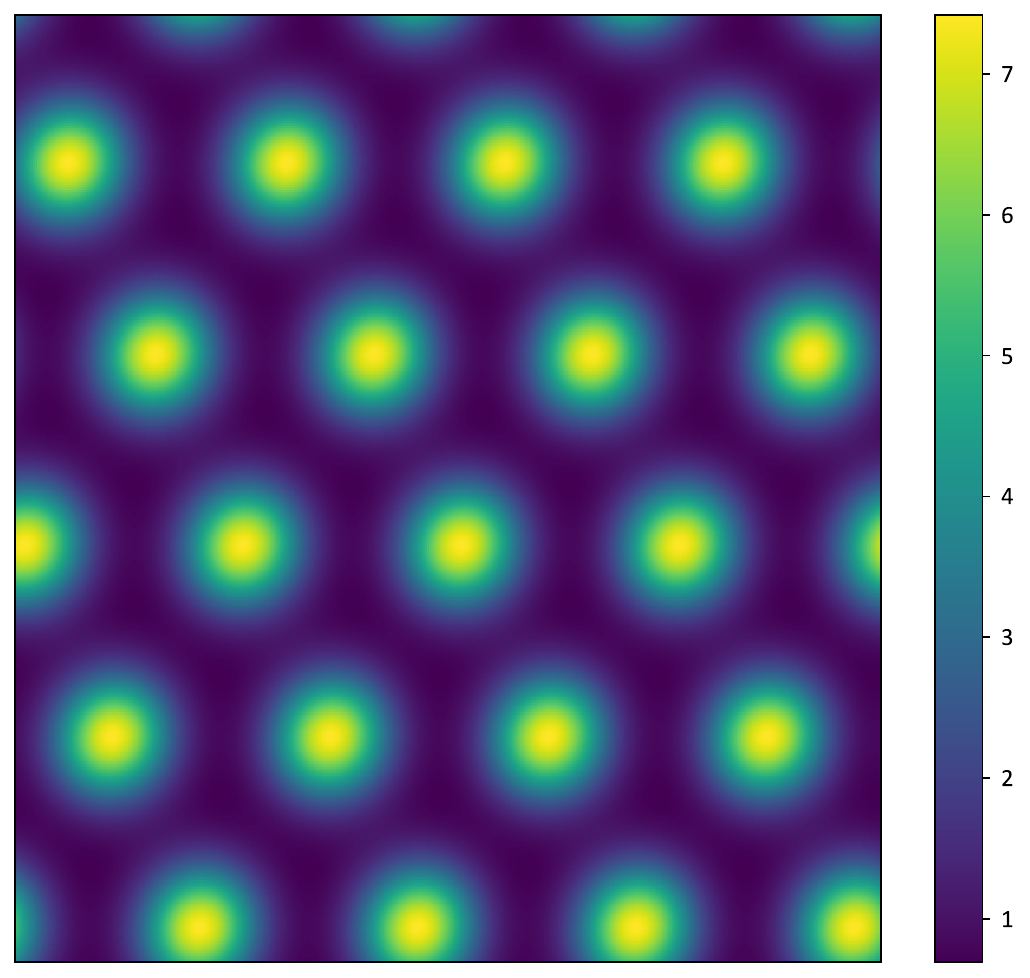}
		\caption{Turing pattern.}
		\label{fig:TP_basal}
	\end{subfigure}
	
	\caption{Dispersion relation and Turing pattern for the system~\eqref{ec:non_linear_basal} with $\alpha_u(u_i,u_o)= \frac{\alpha_i u_i^n}{1+\alpha_o u_o^m}$. The parameters used in the simulation are: $\alpha_i = 16.07$, $\alpha_o = 17.05$, $\beta_{u,u} = 4.68$, $\gamma_u = 4.77$, $m = 4$, $n = 2$,
		$\Gamma_u = 2.37$, $D_{u,i} = 0.01$, $D_{u,o} = 0.25$,
		$u_i^* = 2.16628$, $u_o^* = 0.71906$. As before homogeneous Neumann boundary conditions are imposed; while the domain is a square of side $L = 2$.}
	\label{fig:TP_DR_basal}
\end{figure}

As an illustrative example satisfying these conditions, we consider the Hill-type production function

\begin{equation}
	\alpha_u(u_i, u_o) = \frac{\alpha_i u_i^n}{1 + \alpha_o u_o^m}
	\label{ec:fun_non_lineal_basal}
\end{equation}
where $\alpha_i$ controls the self-activation strength, $\alpha_o$ sets the sensitivity to extracellular inhibition, and $n$, $m$ are cooperativity exponents. The resulting dispersion relation and Turing pattern are shown in Figure~\ref{fig:TP_DR_basal}.

As stated before, single-morphogen Turing patterns have been previously reported in mechanochemical systems \cite{recho_theory_2019} and cell lattice models \cite{wang_periodic_2022}. In addition, Mahashri et al. \cite{mahashri_patterns_2026} showed recently that a single morphogen diffusing across layered two-dimensional media, with nonlinear coupling between layers, is also able to generate stable spatial and temporal patterns. Our framework shares the same core insight — that structural heterogeneity, rather than chemical diversity, can unlock pattern-forming capacity — but differs in that compartmentalization here reflects the biologically grounded separation between intracellular and extracellular domains, rather than geometrically adjacent layers.

Our analysis is conceptually related to the cell lattice model of \cite{wang_periodic_2022}, as the continuous PDE framework can be interpreted as a mean-field description of a cellular lattice; however, in contrast to synthetic lattice models, our framework is derived from first principles \cite{hernandez_mean-field_2026}. The model could also incorporate mechanical features of tissues, such as extracellular fluid velocity \cite{recho_theory_2019} or tissue growth, which we plan to address in future work. Overall, these examples highlight that incorporating tissue structure is essential for studying Turing patterns in biological systems, enabling behaviors that are not accessible in homogeneous models.
	
\section{Conclusions}
\label{sec:conclusions}
	
We have shown that a single molecular species can exhibit Turing instability in a compartmentalized tissue model when either the basal production rate of the morphogen or the membrane transport coupling the intracellular and extracellular fields is nonlinear. The key insight is that compartmentalization structurally doubles the concentration field, introducing the degree of freedom required to satisfy activator–inhibitor conditions even in the absence of a second morphogen. Our results show that tissue microstructure is not merely a passive backdrop for reaction–diffusion processes, but an active determinant of morphogenetic capacity. These findings expands the class of biological systems that could exhibit Turing patterning, suggesting new directions for experimental investigation in single-morphogen signaling contexts.

	\section{Acknowledgments}
	\begin{itemize}
		\item Alejandro Vald{\'e}s L{\'o}pez gratefully acknowledges the financial support provided by the Secretar{\'i}a de Ciencia, Humanidades, Tecnolog{\'i}a e Innovaci{\'o}n (SECIHTI) under Grant No. 4016891.
		
	\end{itemize}

	\section{Declaration of competing interest}
	The authors declare that they have no known competing financial interests or personal relationships that could have appeared to influence the work reported in this paper.
	
	\section{Data availability}
	No data were used for the research described in the article.

	


\begin{thebibliography}{10}
		\expandafter\ifx\csname url\endcsname\relax
		\def\url#1{\texttt{#1}}\fi
		\expandafter\ifx\csname urlprefix\endcsname\relax\def\urlprefix{URL }\fi
		\expandafter\ifx\csname href\endcsname\relax
		\def\href#1#2{#2} \def\path#1{#1}\fi
		
		\bibitem{turing_chemical_1952}
		A.~M. Turing, The chemical basis of morphogenesis, Philosophical Transactions
		of the Royal Society of London. Series B, Biological Sciences 237~(641)
		(1952) 37--72, publisher: Royal Society.
		\newblock \href {https://doi.org/10.1098/rstb.1952.0012}
		{\path{doi:10.1098/rstb.1952.0012}}.
		
		\bibitem{gierer_theory_1972}
		A.~Gierer, H.~Meinhardt, A theory of biological pattern formation, Kybernetik
		12~(1) (1972) 30--39.
		\newblock \href {https://doi.org/10.1007/BF00289234}
		{\path{doi:10.1007/BF00289234}}.
		
		\bibitem{murray_mathematical_1989}
		J.~D. Murray, Mathematical {Biology}, Springer, Berlin, Heidelberg, 1989.
		\newblock \href {https://doi.org/10.1007/978-3-662-08539-4}
		{\path{doi:10.1007/978-3-662-08539-4}}.
		
		\bibitem{kondo_reaction-diffusion_2010}
		S.~Kondo, T.~Miura, Reaction-{Diffusion} {Model} as a {Framework} for
		{Understanding} {Biological} {Pattern} {Formation}, Science 329~(5999) (2010)
		1616--1620, publisher: American Association for the Advancement of Science
		(AAAS).
		\newblock \href {https://doi.org/10.1126/science.1179047}
		{\path{doi:10.1126/science.1179047}}.
		
		\bibitem{Kondo2021Studies}
		S.~Kondo, M.~Watanabe, S.~Miyazawa, Studies of {Turing} pattern formation in
		zebrafish skin, Philosophical Transactions of the Royal Society A:
		Mathematical, Physical and Engineering Sciences 379~(2213) (2021) 20200274,
		publisher: Royal Society.
		\newblock \href {https://doi.org/10.1098/rsta.2020.0274}
		{\path{doi:10.1098/rsta.2020.0274}}.
		
		\bibitem{raspopovic_digit_2014}
		J.~Raspopovic, L.~Marcon, L.~Russo, J.~Sharpe, Digit patterning is controlled
		by a {Bmp}-{Sox9}-{Wnt} {Turing} network modulated by morphogen gradients,
		Science 345~(6196) (2014) 566--570.
		\newblock \href {https://doi.org/10.1126/science.1252960}
		{\path{doi:10.1126/science.1252960}}.
		
		\bibitem{mooney_spatial_1985}
		J.~R. Mooney, B.~N. Nagorcka, Spatial patterns produced by a
		{Reaction}-diffusion system in primary hair follicles, Journal of Theoretical
		Biology 115~(2) (1985) 299--317.
		\newblock \href {https://doi.org/10.1016/S0022-5193(85)80102-8}
		{\path{doi:10.1016/S0022-5193(85)80102-8}}.
		
		\bibitem{sick_wnt_2006}
		S.~Sick, S.~Reinker, J.~Timmer, T.~Schlake,
		\href{https://www.science.org/doi/abs/10.1126/science.1130088}{{WNT} and
			{DKK} {Determine} {Hair} {Follicle} {Spacing} {Through} a
			{Reaction}-{Diffusion} {Mechanism}}, Science 314~(5804) (2006) 1447--1450.
		\newblock \href {https://doi.org/10.1126/science.1130088}
		{\path{doi:10.1126/science.1130088}}.
		\newline\urlprefix\url{https://www.science.org/doi/abs/10.1126/science.1130088}
		
		\bibitem{landge_pattern_2020}
		A.~N. Landge, B.~M. Jordan, X.~Diego, P.~Müller, Pattern formation mechanisms
		of self-organizing reaction-diffusion systems, Developmental Biology 460~(1)
		(2020) 2--11.
		\newblock \href {https://doi.org/10.1016/j.ydbio.2019.10.031}
		{\path{doi:10.1016/j.ydbio.2019.10.031}}.
		
		\bibitem{paul_widespread_2024}
		S.~Paul, J.~Adetunji, T.~Hong, Widespread biochemical reaction networks enable
		{Turing} patterns without imposed feedback, Nature Communications 15~(1)
		(2024) 8380.
		\newblock \href {https://doi.org/10.1038/s41467-024-52591-0}
		{\path{doi:10.1038/s41467-024-52591-0}}.
		
		\bibitem{szili_origin_1997}
		L.~Szili, J.~Tóth, On the origin of {Turing} instability, Journal of
		Mathematical Chemistry 22~(1) (1997) 39--53.
		\newblock \href {https://doi.org/10.1023/A:1019159427561}
		{\path{doi:10.1023/A:1019159427561}}.
		
		\bibitem{mendez_reactiontransport_2010}
		V.~Méndez, S.~Fedotov, W.~Horsthemke, Reaction–{Transport} {Systems},
		Springer {Series} in {Synergetics}, Springer, Berlin, Heidelberg, 2010.
		\newblock \href {https://doi.org/10.1007/978-3-642-11443-4}
		{\path{doi:10.1007/978-3-642-11443-4}}.
		
		\bibitem{henry_fractional_2000}
		B.~I. Henry, S.~L. Wearne, Fractional reaction-diffusion, Physica a-Statistical
		Mechanics and Its Applications 276~(3-4) (2000) 448--455, publisher: Elsevier
		Science.
		\newblock \href {https://doi.org/10.1016/S0378-4371(99)00469-0}
		{\path{doi:10.1016/S0378-4371(99)00469-0}}.
		
		\bibitem{satnoianu_turing_2000}
		R.~A. Satnoianu, M.~Menzinger, P.~K. Maini, Turing instabilities in general
		systems, Journal of Mathematical Biology 41~(6) (2000) 493--512.
		\newblock \href {https://doi.org/10.1007/s002850000056}
		{\path{doi:10.1007/s002850000056}}.
		
		\bibitem{wang_periodic_2022}
		S.~Wang, J.~Garcia-Ojalvo, M.~B. Elowitz, Periodic spatial patterning with a
		single morphogen, Cell Systems 13~(12) (2022) 1033--1047.e7.
		\newblock \href {https://doi.org/10.1016/j.cels.2022.11.001}
		{\path{doi:10.1016/j.cels.2022.11.001}}.
		
		\bibitem{recho_theory_2019}
		P.~Recho, A.~Hallou, E.~Hannezo, Theory of mechanochemical patterning in
		biphasic biological tissues, Proceedings of the National Academy of Sciences
		of the United States of America 116~(12) (2019) 5344--5349.
		\newblock \href {https://doi.org/10.1073/pnas.1813255116}
		{\path{doi:10.1073/pnas.1813255116}}.
		
		\bibitem{mahashri_patterns_2026}
		N.~Mahashri, A.~L. Krause, M.~Chandru, T.~E. Woolley,
		\href{https://arxiv.org/abs/2605.05063v1}{Patterns in {Time} and {Space} from
			a {Single} {Morphogen} via {Nonlinear} {Layering}}, arXiv.org (May 2026).
		\newline\urlprefix\url{https://arxiv.org/abs/2605.05063v1}
		
		\bibitem{keener_mathematical_2009}
		J.~Keener, J.~Sneyd, Mathematical {Physiology}, Vol. 8/1 of Interdisciplinary
		{Applied} {Mathematics}, Springer, New York, NY, 2009.
		\newblock \href {https://doi.org/10.1007/978-0-387-75847-3}
		{\path{doi:10.1007/978-0-387-75847-3}}.
		
		\bibitem{hernandez_mean-field_2026}
		D.~Hernández, Alejandro Valdés López, E.~C. Herrera-Hernández,
		\href{http://arxiv.org/abs/2606.10355}{Mean-field models for morphogenetic
			processes in physiological contexts}, arXiv:2606.10355 [nlin.PS] (Jun. 2026).
		\newblock \href {https://doi.org/10.48550/arXiv.2606.10355}
		{\path{doi:10.48550/arXiv.2606.10355}}.
		\newline\urlprefix\url{http://arxiv.org/abs/2606.10355}
		
		\bibitem{woolley_bespoke_2021}
		T.~E. Woolley, A.~L. Krause, E.~A. Gaffney, Bespoke {Turing} {Systems},
		Bulletin of Mathematical Biology 83~(5) (2021) 41.
		\newblock \href {https://doi.org/10.1007/s11538-021-00870-y}
		{\path{doi:10.1007/s11538-021-00870-y}}.
		
		\bibitem{gudyka_concentration-dependent_2024}
		J.~Gudyka, J.~Ceja-Vega, K.~Ivanchenko, Z.~Morocho, M.~Panella,
		A.~Gamez~Hernandez, C.~Clarke, E.~Perez, S.~Silverberg, S.~Lee,
		Concentration-{Dependent} {Effects} of {Curcumin} on {Membrane}
		{Permeability} and {Structure}, ACS Pharmacology \& Translational Science
		7~(5) (2024) 1546--1556.
		\newblock \href {https://doi.org/10.1021/acsptsci.4c00093}
		{\path{doi:10.1021/acsptsci.4c00093}}.
		
		\bibitem{queralt-martin_ion_2026}
		M.~Queralt-Martín, L.~M. Alvero-González, D.~A. Perini, E.~García-Giménez,
		A.~Alcaraz, Ion transport in biological ion channels beyond classical
		electrostatics. {Nanoscale} confinement, non-linear concentration patterns
		and interfacial effects, Biophysical Reviews (Feb. 2026).
		\newblock \href {https://doi.org/10.1007/s12551-026-01413-2}
		{\path{doi:10.1007/s12551-026-01413-2}}.
		
		\bibitem{kim_permeability_2022}
		W.~K. Kim, S.~Milster, R.~Roa, M.~Kanduč, J.~Dzubiella, Permeability of
		{Polymer} {Membranes} beyond {Linear} {Response}, Macromolecules 55~(16)
		(2022) 7327--7339.
		\newblock \href {https://doi.org/10.1021/acs.macromol.2c00605}
		{\path{doi:10.1021/acs.macromol.2c00605}}.
		
		\bibitem{milster_feedback-controlled_2023}
		S.~Milster, W.~K. Kim, J.~Dzubiella, Feedback-controlled solute transport
		through chemo-responsive polymer membranes, The Journal of Chemical Physics
		158~(10) (2023) 104903.
		\newblock \href {https://doi.org/10.1063/5.0135707}
		{\path{doi:10.1063/5.0135707}}.
		
		\bibitem{zwicker_py-pde_2020}
		D.~Zwicker, py-pde: {A} python package for solving partial differential
		equations, Journal of Open Source Software 5~(48) (2020) 2158.
		\newblock \href {https://doi.org/10.21105/joss.02158}
		{\path{doi:10.21105/joss.02158}}.
		
		\bibitem{maini_turings_2012}
		P.~K. Maini, T.~E. Woolley, R.~E. Baker, E.~A. Gaffney, S.~S. Lee, Turing's
		model for biological pattern formation and the robustness problem, Interface
		Focus 2~(4) (2012) 487--496, publisher: Royal Society.
		\newblock \href {https://doi.org/10.1098/rsfs.2011.0113}
		{\path{doi:10.1098/rsfs.2011.0113}}.
		
		\bibitem{vittadello_turing_2021}
		S.~T. Vittadello, T.~Leyshon, D.~Schnoerr, M.~P.~H. Stumpf,
		\href{https://royalsocietypublishing.org/doi/full/10.1098/rsta.2020.0272}{Turing
			pattern design principles and their robustness}, Philosophical Transactions
		of the Royal Society A: Mathematical, Physical and Engineering Sciences
		379~(2213) (2021) 20200272.
		\newblock \href {https://doi.org/10.1098/rsta.2020.0272}
		{\path{doi:10.1098/rsta.2020.0272}}.
		\newline\urlprefix\url{https://royalsocietypublishing.org/doi/full/10.1098/rsta.2020.0272}
		
		\bibitem{enomoto_autocrine_2023}
		Y.~Enomoto, H.~Katsura, T.~Fujimura, A.~Ogata, S.~Baba, A.~Yamaoka, M.~Kihara,
		T.~Abe, O.~Nishimura, M.~Kadota, D.~Hazama, Y.~Tanaka, Y.~Maniwa, T.~Nagano,
		M.~Morimoto, Autocrine tgf-{$\beta$}-positive feedback in profibrotic
		at2-lineage cells plays a crucial role in non-inflammatory lung fibrogenesis,
		Nature Communications 14~(1) (2023) 4956.
		\newblock \href {https://doi.org/10.1038/s41467-023-40617-y}
		{\path{doi:10.1038/s41467-023-40617-y}}.
		
		\bibitem{waters_quorum_2005}
		C.~M. Waters, B.~L. Bassler, {QUORUM} {SENSING}: {Cell}-to-{Cell}
		{Communication} in {Bacteria}, Annual Review of Cell and Developmental
		Biology 21~(Volume 21, 2005) (2005) 319--346.
		\newblock \href {https://doi.org/10.1146/annurev.cellbio.21.012704.131001}
		{\path{doi:10.1146/annurev.cellbio.21.012704.131001}}.
		
		\bibitem{moreno-gamez_quorum_2023}
		S.~Moreno-Gámez, M.~E. Hochberg, G.~S. van Doorn, Quorum sensing as a
		mechanism to harness the wisdom of the crowds, Nature Communications 14~(1)
		(2023) 3415.
		\newblock \href {https://doi.org/10.1038/s41467-023-37950-7}
		{\path{doi:10.1038/s41467-023-37950-7}}.
		
		\bibitem{kholodenko_cell-signalling_2006}
		B.~N. Kholodenko, Cell-signalling dynamics in time and space, Nature Reviews
		Molecular Cell Biology 7~(3) (2006) 165--176.
		\newblock \href {https://doi.org/10.1038/nrm1838} {\path{doi:10.1038/nrm1838}}.
		
		\bibitem{purvis_encoding_2013}
		J.~E. Purvis, G.~Lahav, Encoding and {Decoding} {Cellular} {Information}
		through {Signaling} {Dynamics}, Cell 152~(5) (2013) 945--956.
		\newblock \href {https://doi.org/10.1016/j.cell.2013.02.005}
		{\path{doi:10.1016/j.cell.2013.02.005}}.
		
		\bibitem{shvartsman_autocrine_2002}
		S.~Y. Shvartsman, M.~P. Hagan, A.~Yacoub, P.~Dent, H.~S. Wiley, D.~A.
		Lauffenburger, Autocrine loops with positive feedback enable
		context-dependent cell signaling, American Journal of Physiology-Cell
		Physiology 282~(3) (2002) C545--C559, publisher: American Physiological
		Society.
		\newblock \href {https://doi.org/10.1152/ajpcell.00260.2001}
		{\path{doi:10.1152/ajpcell.00260.2001}}.
		
	\end{thebibliography}

\end{document}